\documentstyle[prl,aps]{revtex}
\newcommand{\bs}{\bf}
\newcommand{\dz} {{\partial  \over \partial z}}
\newcommand{\dt} {{\partial  \over \partial t}}

\begin{document}

\title{Evaporative cooling of an atomic beam}

\author{E. Mandonnet, A. Minguzzi$^*$, R. Dum,
 I. Carusotto$^*$,\\ Y. Castin, and J. Dalibard}

\address{Laboratoire Kastler Brossel, 
24 rue Lhomond, 75005 Paris, France\\
$^*$ INFM, Scuola Normale Superiore, Piazza dei Cavalieri 7, I-56126 Pisa, Italy}

\maketitle

\begin{abstract}
We present a theoretical analysis of 
the evaporative cooling of an atomic beam 
propagating in a magnetic guide. 
Cooling is provided by transverse
evaporation.
The atomic dynamics inside the guide is analyzed by solving the
Boltzmann equation with two different approaches: an approximate
analytical ansatz and a Monte-Carlo simulation.
Within their domain of validity, these
two methods are found to be in very good agreement
with each other.
They allow us to determine how
the phase-space density and the flux of the beam vary along
its direction of propagation.
We find a significant increase for the phase-space density 
along the guide for realistic experimental parameters. 
By extrapolation, we estimate the length of the
beam needed to reach quantum degeneracy.

\end{abstract}

%\pacs{03.75.Fi,05.30.Jp}

\section{Introduction}

Forced evaporative cooling of trapped 
gases is a very powerful technique to 
increase the phase-space density of an 
ensemble of atoms up to quantum degeneracy \cite{Ketterle97}.
Particles with a 
sufficiently large energy (typically 5 times 
the thermal energy $k_BT$) are eliminated. 
Elastic collisions between the remaining 
particles restore thermal equilibrium with a lower temperature 
and an increased phase-space density. 
The most prominent success of evaporative 
cooling is the achievement of Bose-Einstein condensation 
of alkali and hydrogen atomic 
gases\cite{Anderson95,Bradley957,Davis95,Fried98}. 

In this paper we present
another possible application of evaporative 
cooling.
We consider atoms moving freely along 
the $z$-axis and transversely confined 
by a magnetic field gradient, which 
provides a harmonic potential in the 
$x-y$ plane (see fig. \ref{SHEME}). 
The atoms are injected in 
the plane $z=0$ with  given flux, longitudinal 
and transverse velocity and spatial distributions. 
We suppose that a radio-frequency field 
is applied 
so that only atoms inside a tube-shaped region remain trapped, those outside
are evaporated.
The bias magnetic field along the $z$-direction is 
adjusted so that the cross-section of the tube decreases 
as $z$ increases.
As the atoms move forward in
the magnetic guide, the ones with a large
transverse energy are evaporated. We then 
rely on elastic collisions between the
remaining atoms to decrease the longitudinal 
velocity width and to increase the 
phase-space density of the beam. 

We investigate theoretically the efficiency
of this evaporative cooling scheme by solving the Boltzmann
equation with two different approaches:
one is based on a truncated distribution ansatz for the phase-space
distribution function, while the other one
is a direct Monte-Carlo simulation of the atomic dynamics 
inside the guide.
For a suitable geometry of the evaporation scheme, the
results of the analytical
method fit quite well those of the Monte-Carlo simulation.
They show that, with reasonable experimental parameters,
the cooling process can lead to a spectacular increase of the 
phase-space density of the atomic beam.
This might be considered as a possible realization of a continuous
atom laser \cite{atom_laser,manip}. Such a coherent source of atoms
would have fascinating applications in atom interferometry and holography, 
metrology and atomic clocks, and nonlinear atom optics \cite{Helmerson}.

The paper is organized as follows: 
in section \ref{MODEL} we describe the magnetic guide
(Sec.\ref{GUIDE}), 
we give some typical parameters of the atomic beam source (Sec.\ref{PARA}),
we introduce the Boltzmann equation (Sec.\ref{COLL}),
and we detail the evaporation scheme (Sec.\ref{EVAP}) ; in section 
\ref{APPROX} we use a truncated Boltzmann distribution to obtain
an approximate analytical solution of  
the Boltzmann equation for the case of a 1D evaporation scheme ;
in section \ref{RES} we explain the Monte-Carlo simulation
and we analyze the results of the two methods in section \ref{sec_disc}.
The paper is concluded in section \ref{sec_per} with a brief discussion
of the coherence properties of the beam
when the cooling is sufficiently  efficient to reach quantum
degeneracy, that is when the  spatial density is of the order
of $\lambda^{-3}$, where $\lambda$ is the local thermal wavelength.

\section{Model considered}\label{MODEL}

\subsection{The atomic guide}\label{GUIDE}

We assume that the magnetic guide
consists of four parallel wires carrying
the same current $\pm I$ along the $z$
 direction (see fig. \ref{SHEME}); each
wire is  
at a distance $a$ from the line $x=y=0$. 
The resulting magnetic field is 
${\bs B} = b'(x,-y,0)$, with $b'=2\mu_0 I/(\pi a^2)$.
We superimpose a longitudinal magnetic field 
${\bs B_0}$ along the $z$ axis 
so that the modulus of the total magnetic field 
can be written for $x,y$ sufficiently
small:
\begin{equation}
B({\bs r})=\left(B_0^2+b'^2(x^2+y^2)\right)^{1/2}
\simeq B_0 + \frac{{b'}^2}{2B_0}(x^2 + y^2)
\quad .
\label{FIELD}
\end{equation}
A magnetic moment $\mu$ which is prepared in the 
direction opposite to the local ${\bs B}$ will 
therefore experience a trapping harmonic 
potential in the transverse directions.
This potential is necessarily isotropic in the $x-y$ 
plane because of Maxwell equations
for magnetostatics. The oscillation frequency 
in this plane is given by
\begin{equation}
\Omega_\bot=(\mu {b'}^2 /(m B_0))^{1/2}
\end{equation}
where $m$ is the atomic mass.

Typical experimental values are $a=4$~mm,
 $I=500$~A, and $B_0=1$~mT. 
For rubidium atoms with $\mu$ equal to the 
Bohr magneton, this leads to an oscillation 
frequency $\Omega_\bot \sim 2\pi \times 1000$~Hz. 
The quadratic expansion leading 
to (\ref{FIELD}) is valid for transverse
 temperatures up to $500\ \mu$K.

\subsection{Parameters of the atomic beam source} \label{PARA}

In our model the atoms - for the envisaged experiment we 
consider rubidium atoms -
enter the magnetic guide in the 
plane $z=0$ with 
a Gaussian velocity and space distribution with 
the same initial temperature $T_0$ for transverse and longitudinal degrees of freedom.
The longitudinal velocity distribution is
centered around a non zero value
$\bar v_0$.
The ratio $\bar v_0/ \Delta v_0$ ,
where $\Delta v_0=\sqrt{k_B T /m}$ is the {\it r.m.s.} 
of the Gaussian velocity distribution,
should be larger than 1 to  ensure
that only a very small fraction of atoms is initially 
moving ``upstream", 
that is with a negative longitudinal velocity. 
We choose  in the following
\begin{equation}
\bar v_0/ \Delta v_0 =3. 
\end{equation}
For an initial temperature $T_0=$400~$\mu$K, this gives
$\bar v_0=60$~cm/s. 
The {\it r.m.s.}
of the spatial transverse distribution is $ R_{{\bot}0}
=\Delta v_0/\Omega_\bot$.

We further assume an initial on-axis density $n_0$ at $x=y=0$ of the atomic beam of
 $8\times 10^{11}$~cm$^{-3}$
which corresponds to a flux 
$\Phi_0= 2\pi R_{{\bot}0}^2 n_0 {\bar v}_0 \simeq 3 \times 10^{9}$~s$^{-1}$
\cite{flux}.
The initial stage of the evaporative cooling 
within the guide can be described by
classical dynamics as
$n_0\;\lambda_0^{3} \approx 7 \times 10^{-7} \ll 1$, 
where $\lambda_0=h/\sqrt{2\pi m k_B T_0}$ is the 
initial thermal wavelength of the gas.

\subsection{ Collisional dynamics inside the guide}\label{COLL}

The collisional dynamics inside the guide can be theoretically
described by the Boltzmann equation, which gives the time evolution
of the atomic phase-space density $f({\bs r},{\bs p},t)$:
\begin{equation} \label{BOLTZ}
\frac{\partial f}{\partial t}+ \frac{\bs p}{m}\cdot {\bs \nabla}_{\bs r}f
-{\bs \nabla}_{\bs r} U  \cdot {\bs \nabla}_{\bs p}f= I_{\rm coll}\left[
f \right]\quad ,
\end{equation}
where $U(x,y,z)=U_x(x)+U_y(y)$ with 
$U_x(x)=\frac{1}{2} m \Omega_{\bot}^2 x^2$ and
$U_y(y)=\frac{1}{2} m \Omega_{\bot}^2 y^2$.
The distribution $f$ is normalized so that its integral over
momentum gives the spatial density. Considering only elastic 
and isotropic collisions 
between guided atoms, we write the collisional integral as:  
\begin{eqnarray}
I_{\rm coll}\left[ f \right]&=& \frac{\sigma}{ \pi m^2 }
\int d^3 p_2 \int d^3 p_3 \int d^3 p_4 
\; \left(f({\bf r}, {\bf p}_3) f({\bf r}, {\bf p}_4)
- f({\bf r}, {\bf p}) f({\bf r}, {\bf p}_2) \right)
\nonumber \\
&& \delta ({\bs p} + {\bs p}_2 - {\bs p}_3 - {\bs p}_4)
\; \delta \left(\frac{p^2}{2m}+\frac{p_2^2}{2m}- \frac{p_3^2}{2m}- \frac{p_4^2}{2m}
\right).
\label{TC}
\end{eqnarray}
Here we assumed that the cross-section $\sigma$ is independent of atomic momentum,
a valid assumption for alkali atoms (in the absence of a zero energy
resonance) if the temperature
is low enough so that collisions essentially occur in the $s$-wave regime. 
For instance for polarized rubidium atoms, 
the region of $s$-wave scattering
extends up to $400\ \mu$K, corresponding to a {\it r.m.s.} velocity 
$\Delta v_0=20$~cm/s. Above this value, 
$d$-wave scattering is not negligible,
and it may significantly modify the  results of this paper. 
We recall that $p$-wave scattering 
(more generally any odd wave scattering) is 
forbidden at any energy for polarized bosons.

The collision rate of the atomic beam source 
$\gamma_{\rm coll}$ is related to the on-axis density $n_0$
by $\gamma_{\rm coll}=(2/\sqrt{\pi})\, n_0\, \sigma \, \Delta v_0$.
For the $s$-wave collisional cross-section of rubidium 
($\sigma=7.6 \times 10^{-16}$~m$^2$) we get: 
\begin{equation}
\gamma_{\rm coll}/\Omega_\bot=0.02\quad, 
\end{equation}
which means that an atom performs on average several transverse 
oscillations between two collisions, that is we are in the collisionless
regime for the transverse degrees of freedom.

\subsection{Evaporation scheme}\label{EVAP}

Evaporation along the beam eliminates 
atoms outside a section in the transverse $xy$ plane. In order
to optimize the efficiency of this evaporation, we assume that the 
size of this section varies with $z$ in a controlled way (forced evaporation).
In practice this can be achieved using
a radio-frequency field at a fixed frequency $\nu_{\rm r.f.}$, and
a spatially varying bias field $B_0$ and gradient $b'$, producing thus 
a spatially varying bottom of the 
magnetic well while keeping a constant $\Omega_\bot\propto b'/\sqrt{B_0}$ \cite{gravity}. 
The stability constraints concern mostly the bias field $B_0$ as it controls directly
the effective local trapping potential depth $\Delta U_0 = (h\nu_{\rm r.f.}/g)
-\mu B_0$ where $g$ is the Land\'e factor of the atomic level. 
As we shall see in section \S \ref{approche} a typical value for
$\Delta U_0$ at the point where quantum degeneracy is reached is  $\sim 1\;\mu$K.
This value is comparable to the potential depth typically used in
Bose-Einstein condensation experiments, so that we do not anticipate any
particular difficulty with the control of $B_0$.
Another possibility to achieve evaporation
is to place an absorbing material at a controlled distance 
$\Lambda_{\rm evap}(z)$ from the center of the guide.

Evaporation is taken into account in the model by putting
$f({\bs r}, {\bs p}, t)=0$ if the phase-space cell 
$\{{\bs r}, {\bs p}\}$ is outside the domain where atoms are trapped. 
For the case of a radio-frequency evaporation, which is cylindrically symmetric as 
long as gravity is negligible, the evaporation criterion is: 
\begin{equation} \label{CUT2D}
x^2+y^2 > \Lambda_{\rm evap}^2(z)\quad
\end{equation}
where $\Lambda_{\rm evap}(z)$ is determined by the radio-frequency field:
$g\mu (B_0^2+b'^2 \Lambda_{\rm evap}^2)^{1/2}=h\nu_{\rm rf}$.
This corresponds to a 2D evaporation
scheme. We also will consider in this paper a 1D scheme with  the criterion
\begin{equation} \label{CUT1D}
x^2 > \Lambda_{\rm evap}^2(z) \quad.
\end{equation}
As we will see below, this 1D scheme allows an approximate analytical treatment.
We also define the cut-off energy:
\begin{equation}
\epsilon_c(z)= {1 \over 2} m \Omega_\bot^2 \Lambda_{\rm evap}^2(z)  
\quad.
\end{equation}

\section{1D hydrodynamic approach}\label{APPROX}

\subsection{The truncated Boltzmann distribution ansatz}

In this section we restrict the discussion to the 1D scheme of Eq. (\ref{CUT1D}).
We now assume that the flow of the gas is in the hydrodynamic regime
along the longitudinal direction, that is the macroscopic quantities
 - such as density
or mean velocity  -
vary slowly with $z$ on a scale given by the mean free path 
$d(z)=\Delta v(z)/\gamma_{\rm coll}(z)$.
The assumption of a hydrodynamic regime implies a local thermodynamic equilibrium
characterized by a local temperature $T(z)$ at abscissa $z$.

We also assume that  
the mean free path is much larger
than the transverse extension $R_\bot(z)$ of the beam, or equivalently 
$\gamma_{\rm coll}(z) \ll \Omega_\bot$.
As a consequence, in the 1D evaporation scheme, if an atom 
emerges from a collision with a kinetic+potential energy 
along the $x$-axis higher than the cut-off 
$\epsilon_c(z)$, it will fulfil   
after the further oscillation -- and before it undergoes
another collision -- the condition (\ref{CUT1D})
and it will be evaporated.
Therefore it is equivalent to formulate the evaporation criterion 
either in terms of the coordinate $x$ or in terms of  
$\epsilon_x(x,p_x)= U_{x}(x)+ {p_{x}^2}/{(2m)}$.
Consequently we replace
Eq.(\ref{CUT1D}) by
\begin{equation} \label{CUTE}
{1 \over 2} m \Omega_\bot^2 x^2  + {p_x^2 \over 2 m} > \epsilon_c(z)
\quad .
\end{equation}

The assumption of a local thermodynamic equilibrium of the 
gas at each abscissa $z$ around
a mean velocity ${\bar v}(z)={\bar p}(z)/m$ suggests the 
following ansatz for the classical phase-space distribution
\cite{amsterdam1}:
\begin{equation} \label{ANS}
f({\bs r},{\bs p})=f_{0}(z)\; e^{-\beta(z) (\epsilon_x+\epsilon_y)}\;
e^{-\beta(z)(p_z-{\bar p}(z))^2/2m}\; Y(\epsilon_c(z)-\epsilon_x(x,p_x)).
\end{equation}
Here $\beta(z)=1/(k_B T(z))$
and $Y$ is the Heaviside step function.
The truncation takes into account the evaporation:
$f=0$ for an atom whose
energy in the transverse direction $x$ exceeds 
the local depth $\epsilon_c(z)$. 
The on-axis phase-space density $f_0(z)$ is calculated by the 
normalization condition 
$\int dx dy \int d^3p \; f =\rho_{\rm lin}(z) $;
here $\rho_{\rm lin}(z)$ is the linear density of the gas, that is 
the number of particles per unit 
length in the guide.
A straightforward calculation leads to 
\begin{equation}
f_{0}(z)=\frac{1}{1-e^{-\eta(z)}} \; \frac{1}{(2\pi)^{5/2}}\;
 \frac{\rho_{\rm lin}(z)}{(m \Delta v(z))^3 
R_{\bot}^2(z)}
\label{NORM}
\end{equation}
where we have introduced 
the thermal velocity $\Delta v(z)=\sqrt{k_B T(z)/m}$,
the thermal transverse size of the beam
 $R_{\bot}(z)=\sqrt{k_B T(z)/m \Omega_\bot^2}$,
and the quantity 
\begin{equation} \label{ETA}
\eta(z)=\epsilon_c(z)/ k_B T(z),
\end{equation}
which is a crucial parameter to control
the efficiency of evaporation.

This reformulation of the evaporation criterion in the energy domain
is difficult to extend to the 2D scheme. It would require the further
assumption of ergodicity of the transverse motion, which 
is not correct for the present 
{\em axi-symmetric} potential since 
the knowledge of the total transverse energy $\epsilon_\bot=\epsilon_x+\epsilon_y$
is not a 
sufficient criterion to determine whether an atomic trajectory will
be evaporated. For instance atoms moving along transverse linear trajectories are
evaporated when $\epsilon_\bot>\epsilon_c(z)$, while 
atoms with transverse circular trajectories remain
trapped as long as $\epsilon_\bot$ is below $2\, \epsilon_c(z)$.
This 2D case will be treated using a Monte-Carlo
technique in section \ref{2D1D} \cite{remarque}.

\subsection{Hydrodynamic equations}

We now derive
a closed set of partial differential equations for the three
macroscopic quantities $T(z,t)$, $\rho_{\rm lin}(z,t)$, and ${\bar p}(z,t)$.
We note that $f({\bs r},{\bs p})$  in the
transverse direction depends only on $\epsilon_x$ 
and $\epsilon_y$, quantities preserved by 
the Hamiltonian evolution. Therefore Eq.(\ref{BOLTZ})
reduces to
\begin{equation} \label{BOLTZ2}
\frac{\partial f}{\partial t}+ \frac{p_z}{m} \frac{\partial f}{\partial z}
= I_{\rm coll}\left[
f \right]\quad .
\end{equation}
We now proceed in a way analogous to the derivation of the
standard hydrodynamic equations from the Boltzmann equation:
multiplying Eq.(\ref{BOLTZ2}) by $1$, $p_z$, and $\epsilon_x+\epsilon_y+
p_z^2/2m$, respectively,  and integrating over the five variables $x$,$y$,$p_x$,
$p_y$,$p_z$ gives:
\begin{eqnarray}
  \dt  \rho_{\rm lin}  +
    \frac{\partial}{\partial z}\left [ {{\bar p} \over m}\,
    \rho_{\rm lin}\right]&=&
	-\,{\Gamma}\, \rho_{\rm lin} \label{HY1}
 \\
   \frac{\partial}{\partial t} \left[  {\bar p} \,\rho_{\rm lin}\right]
    +\frac{\partial}{\partial z}
     \left[ \left(k_B T +\frac{{\bar p}^2}{m}\right)\,
    \rho_{\rm lin} \right]& =& 
-\,{\Gamma}\,{\bar p} \, \rho_{\rm lin} \label{HY2} 
\\
   \frac{\partial}{\partial t}\left[  \left(
      {\bar \epsilon}_x +
    \frac{3}{2}k_B T
    +\frac{{\bar p}^2}{2m} \right)\,\rho_{\rm lin} \right] +
    \frac{\partial}{\partial z}\left[  \left(
{\bar \epsilon}_x +
    \frac{5}{2} k_B T +
    \frac{{\bar p}^2}{2m} \right)\, \frac{{\bar p}}{m}\,
\rho_{\rm lin} \right]& =&
    -\,\left(  {\Gamma}\;{{\bar p}^2 \over 2m} +
    {\Gamma}_{\rm \epsilon}\;k_BT\right) \;  \rho_{\rm lin}
\label{HY3}
\end{eqnarray}
where 
\begin{equation}
{\bar \epsilon}_{x}= 
\frac{1}{\rho_{\rm lin}} \int dxdy \int d^3p \; \epsilon_{x} f({\bs r},{\bs p})
=k_B T \left( 1 - \frac{\eta}{e^{\eta}-1} \right) .
\end{equation}
Equations (\ref{HY1},\ref{HY2},\ref{HY3}) are the equations of conservation for 
the number, the momentum, and the energy of the particles.
On the right-hand side we have source terms due to evaporation:
${\Gamma}$ is the evaporation rate at abscissa $z$, 
and ${\Gamma}_{\rm \epsilon}\,k_BT$ is the 
rate for the decrease of energy  in the local 
reference frame moving at velocity ${\bar v}(z)$. 

The 1D evaporation model \cite{amsterdam2} allows to  derive explicit expressions 
for ${\Gamma}$ and ${\Gamma}_{\rm \epsilon}$.
We obtain for the loss rate of particles: 
\begin{equation}
{\Gamma}(z)=\sigma\,\rho_{\rm lin}(z)\,\frac{\Delta v(z)}{R_{\bot}^2(z)}
\, e^{-\eta(z)}\, S(\eta(z)) 
\label{LOSS1}
\end{equation}
The analytical expression of the positive dimensionless coefficient $S(\eta)$ 
is given in the appendix. As shown in fig. \ref{fig:S},
it is a nearly constant quantity, of the order of 0.075, 
when $\eta$ varies between 2 and 10.
Eq.(\ref{LOSS1}) shows that the decay rate $\Gamma$ of the linear density,
given in eq.(\ref{HY1}),
 is proportional to the collisional cross section $\sigma$ 
and  to the local atomic density, as expected for binary 
collisions.

For the loss rate of energy,  we find: 
\begin{equation}
{\Gamma}_{\rm \epsilon}(z) = {\Gamma}(z)\;
\left( \eta(z) + \frac{3}{2} + \tilde{S}(\eta(z)) \right)\quad.
\label{LOSS2}
\end{equation}
The expression of the positive dimensionless coefficient $\tilde{S}(\eta)$ 
is also derived in the appendix, and plotted in fig \ref{fig:S}. It
increases from  0.43 to  0.66 when $\eta$ varies from 2 to 10.

On the left-hand sides of equations (\ref{HY1},\ref{HY2},\ref{HY3}) 
we have neglected the terms arising from the $z$-dependence of the
cut-off $\epsilon_{c}(z)$. These terms would account for 
spilling, that is the loss of particles even in absence of collisions
due to the lowering of the cut-off $\epsilon_{c}(z)$ along the beam. 
Neglection of spilling is valid for $\eta \gg 1$, which is well
verified for the optimal evaporation with realistic initial parameters;
for instance, for the simulations presented in section
\ref{RES}, we have chosen $\eta \simeq 5$ \cite{SPIL}.
For consistency we also replace  in the following
${\bar \epsilon}_{x}$  and $f_0$ by their values for
$\eta \rightarrow \infty$:
\begin{eqnarray}
{\bar \epsilon}_{x} &=& k_B T \\
f_{0}(z)&=&\frac{1}{(2\pi)^{5/2}}\,
 \frac{\rho_{\rm lin}(z)}{(m \Delta v(z))^3 
R_{\bot}^2(z)}
\end{eqnarray}
We finally obtain from Eqs. (\ref{HY1},\ref{HY2},\ref{HY3})  
in the limit of large $\eta$:
\begin{eqnarray}
\left(\dt+\frac{{\bar p}}{m}\dz\right)\rho_{\rm lin}
+\rho_{\rm lin} \frac{\partial}{\partial z}\left [{ {\bar p} \over m}
\right]& =&
     -{\Gamma}\rho_{\rm lin}
  \\
\left(\dt+\frac{{\bar p}}{m}\dz\right)
{\bar p}+\frac{1}{\rho_{\rm lin}}\dz[
k_B T\;\rho_{\rm lin} ]& =&0  
\\
\left(\dt+\frac{{\bar p}}{m}\dz\right )k_B T +
\frac{2}{5}k_B T \dz\left[\frac{{\bar p}}{m}\right]
&=&
k_B T \, (\Gamma-\frac{2}{5}{\Gamma}_{\rm \epsilon})
\end{eqnarray}

\subsection{Stationary regime: $z$-dependence of phase space density}
We are mainly interested in a stationary 
regime  for which  we obtain a set of non-linear equations which can be solved by
standard numerical methods:
\begin{eqnarray}
\dz \left( \rho_{\rm lin}\right) &=& -\frac{m}{{\bar p}} \rho_{\rm lin}
\left( \Gamma \;+\; \Gamma_{\rm \epsilon }\;
\frac{k_B T}{E} \right) 
\label{HYnum1}
\\
\dz \left( \frac{{\bar p}}{m}\right)&=& {\Gamma}_{\rm \epsilon}\;
\frac{k_B T}{E} 
\label{HYnum2} \\
\dz(k_B T)&=&- \frac{mk_BT}{{\bar p}} 
\left(\frac{2}{5}\Gamma_\epsilon -\Gamma + \frac{2}{5}\Gamma_\epsilon  
\frac{k_B T}{E}
\right)
\label{HYnum3}
\end{eqnarray}
where we have put
$$
E\equiv \frac{5}{2}\frac{{\bar p}^2}{m}
-\frac{7}{2}k_B T.
$$ 
\par From the above we see that if ${\bar p}^2/2m$ 
is bigger than $(7/10) k_B T$
({\it i.e.} $E>0$), 
which is indeed the parameter range studied in this article,
the mean velocity increases as a function of $z$.
An interpretation of this result will be given in section \ref{approche}.

A figure of merit for our scheme is the degree of increase of the
phase-space density along the beam. 
As a measure of this increase we take the on-axis phase-space density 
$f_0(z) \propto {\rho_{\rm lin}}/{(k_B T)^{5/2}}$: 
\begin{equation}
\dz \left[ \ln\frac{\rho_{\rm lin}}{(k_B T)^{5/2}} \right]
=\left(
\eta+\tilde{S}(\eta) -2 \right)\; \frac{m}{{\bar p}}\; {\Gamma} .
\end{equation}
Since the quantity $\tilde{S}(\eta)$ is positive (see fig. \ref{fig:S}), 
we conclude that the phase-space density increases  when
$\eta > 2$. 

\section{Monte-Carlo simulations}\label{RES}

In this section we compare results from the approximate analytical ansatz of the previous
section with results of a Monte-Carlo simulation. 
The fact that the Monte-Carlo simulation requires a long
computing time restricts the 
parameter space which can be explored. In particular, as 
explicited below, the length of the
guide influences the memory requirement of the Monte-Carlo simulation.
The total length $L$ of the system, expressed in units of 
the mean free path $d_0=\sqrt{\pi}/(2n_0 \sigma)$
is chosen in the following as 
\begin{equation}
L/d_0= 2500.
\end{equation}
For the parameters given in section \ref{PARA} 
this corresponds to a length
of  3.7~meters;
the average time $T=L/\bar v_0$ for an atom to travel 
from the entrance to the exit
of the guide is  $T=4 \times 10^{4}\; \Omega_\bot^{-1}$ (in absence of collisions
and evaporation);
this time corresponds to  $830\; \gamma_{\rm coll}^{-1}$.

\subsection{Principle of the Monte-Carlo method}
This method has originally been introduced in the context
of molecular dynamics \cite{BIRD}. For the case of dilute gases,
it relies on the idea that one can separate the description of the
collision from that of the motion, allowing a simulation 
of the dynamics on a time 
scale shorter than the mean time between two collisions. 

In essence the approach consists in solving (\ref{BOLTZ}) numerically  by
evolving macro-atoms, each of which representing $\ell$ real atoms.
The macro-atoms evolve in the same potential as the real atoms, and
they have the same initial velocity and position distributions.
Their collisional cross-section is $\ell\sigma$ and their initial
spatial density is $n_0/\ell$, so that the collision rate 
$\gamma_{\rm coll}$
and the two dimensionless parameters ${\bar v}_0/\Delta v_0$ 
and $L/d_0$ are not changed.
This approach is valid since the Boltzmann equation is 
invariant under the scaling $\sigma \rightarrow \ell \sigma$,
$f \rightarrow f/\ell$. 
Using the symmetries of the problem we restrict the evolution
to the first quarter ($x>0$, $y>0$) of the entrance plane $z=0$
reducing the memory requirement by a factor 4.
We evolve the macro-atoms in this first quarter
with reflecting walls at the planes $x=0$ and $y=0$.

We inject on average 84 macro-atoms (21 in the first quarter)
every $\Omega_\bot^{-1}$.
We take  $\ell=5600$ to match the flux
of the atomic beam source presented in section \ref{PARA}. 
In absence of evaporation, 
$N=21 \times 4\times 10^4= 8.4 \times 10^5$
macro-atoms are present on average at a given time since, 
as stated above, it takes on
average $4\times 10^4\, \Omega_\bot^{-1}$ for a particle 
to travel along the guide. 

Binary elastic collisions are taken into account using
a boxing technique\cite{wu96,wu97,dgo98}. We introduce in
position space a 
lattice with a unit cell volume $\delta V$, 
chosen such that the average
occupation 
$p_{\rm occ}$ of any cell is much smaller than 1. 
Collisions occur only
between two macro-atoms occupying the same cell, and  
the time step $\delta t$ is adjusted in such 
a way that the probability $p_{\rm coll}$ of a collisional event during 
$\delta t$ is also much smaller than 1. 
We choose typically  $p_{\rm occ}\sim p_{\rm coll}\sim  10\%$.

Evaporation is implemented in the simulation 
by eliminating the macro-atoms whose coordinates fulfil the chosen 
condition of evaporation. In particular, we have treated both the 
case of evaporation with a 1D position cut (Eq. (\ref{CUT1D})) in order
to compare with the results of the analytical ansatz, and the
case of evaporation with a 2D position cut (Eq. (\ref{CUT2D})).

We let the simulation evolve until
a steady-state is reached. The corresponding
time is $\sim 2\, L/\bar{v}_0 = 8 \times 10^4 \, \Omega_\bot^{-1}$. 
This allows to obtain the average energies along each axis
and the linear density at a given location $z$. From these 
quantities we can predict the decrease in temperature,
the loss of particles and in consequence 
the phase-space density increase.

\subsection{Comparison with the hydrodynamic approach for 1D evaporation}\label{COMP}

The results of the Monte-Carlo and of the hydrodynamic approaches for 1D evaporation 
are plotted in 
Figs. \ref{VELO},\ref{DATAa} and \ref{DATAb}, giving 
the variations with $z$ of  the mean velocity $\bar v(z)$,
the flux $\Phi(z)$, and the phase-space 
density $f_0(z)$.
This set of data has been obtained by choosing
an evaporation barrier $\epsilon_c(z)$
approximately 5 times larger than the mean local 
transverse energy, that is we kept $\eta \simeq 5$; 
as shown below, this value for $\eta$ leads to a gain
of 7 orders of magnitude for the phase space density with
a minimal length requirement. 

Using the Monte-Carlo simulation, we have found that
the three components of the velocities have a nearly Gaussian distribution,
with dispersions equal to within a few percent.
This is consistent with the hypothesis of local 
thermodynamic equilibrium at the basis of the hydrodynamic approach.

The results shown in figs. \ref{VELO}, \ref{DATAa} and  \ref{DATAb}
show an excellent agreement between the
hydrodynamic approach and the Monte-Carlo simulation. 
This allows one to make all optimization
and design procedures for the choice of initial parameters and spatial
variations of the cut using the approximate hydrodynamic treatment while
keeping the Monte-Carlo simulation -- which requires several days
of computation on a work station -- for final checks.

\section{Discussion of the results} \label{sec_disc}

\subsection{1D evaporation}
\label{approche}

Using the equations (\ref{HYnum1}-\ref{HYnum3}) 
in the regime of evaporation with a constant $\eta$, 
we have first determined the optimal choice for $\eta$. As shown in fig. \ref{eta},
the shortest distance providing a $10^7$ gain in phase space density -- required 
in order to reach quantum degeneracy for the initial conditions considered in \S 2-- 
is obtained for 
$\eta \sim 5$. If smaller phase space gain are needed, smaller values of $\eta$ 
would be more appropriate since they would lead to shorter cooling lengths.

We now consider fig.\ref{VELO}, obtained for $\eta=5$, which 
shows a slight increase of the mean velocity
$\bar v(z)$,
as the atoms progress within the guide. This increase can be understood
from kinematic arguments: the cooling along $z$
can be seen very crudely
as a process in which a fast and a 
slow particle collide, one of them being eliminated. Now, at a given location,
the fast particles are renewed with a larger rate than the slow ones,
because it takes less time for them to go from the entrance plane
to the considered location. Therefore, the mean velocity in a location
$z>0$, where particles have already undergone in average several collisions,
is larger than in $z=0$. This acceleration effect becomes negligible as soon as
$k_B T(z) \ll \bar p^2/(2m)$ since the beam is then quasi mono-kinetic.

The figs. \ref{DATAa},\ref{DATAb}
show the flux and the phase-space density as a function of $z$
for $\eta=5$.
A very significant increase of phase-space 
density by a factor 500
proves the efficiency of the evaporative cooling. This increase is
accompanied with a reduction of the flux by a factor 5.5. 
The phase-space density and the flux both vary 
quasi-exponentially with the position $z$.

To get a better understanding of the variations with $z$ of these quantities,
we now consider Eqs. (\ref{HYnum1}-\ref{HYnum3}) in the limiting case where
$\bar v$ is constant and $k_BT\ll E$. Assuming also a constant $\eta$, the solutions of 
these equations are:
\begin{equation}
\rho_{\rm lin}(z) = \rho_{\rm lin}(0)\; \left(1 - \frac{z}{z_c} \right)^{\alpha_c}
\qquad \qquad 
T(z) = T(0)\; \left(1 - \frac{z}{z_c} \right)^{\beta_c}
\end{equation}
with
\begin{equation}
z_c=d_0\, \frac{\bar v}{\Delta v_0}\; \frac{5\, e^\eta}{\pi^{3/2}S(\eta)(\eta + \tilde{S}(\eta)-6 ) }
\qquad\qquad
\alpha_c=\frac{5}{\eta + \tilde{S}(\eta)-6}
\qquad\qquad
\beta_c=\frac{2(\eta+\tilde{S}(\eta)-1 )}{\eta + \tilde{S}(\eta)-6}
\end{equation}

For the parameters of figs.\ref{VELO},\ref{DATAa} and \ref{DATAb},  where $\eta=5$,
we find $\alpha_c\simeq -12,\beta_c\simeq -21.9$ and $z_c\simeq -4300\; d_0
\bar v/\Delta v_0 \simeq -13\,000\; d_0$, so that  the
relevant range of lengths $z$ are much smaller than $|z_c|$. 
Consequently we can approximate the previous results with:
\begin{equation}
\rho_{\rm lin}(z) \simeq \rho_{\rm lin}(0)\; \exp(- \alpha_c z/z_c) 
\qquad \qquad 
T(z) \simeq T(0)\;\exp(-\beta_c z /z_c)
\end{equation}
This leads to exponential variations $\exp(z/z_{\rm p.s.})$ 
and $\exp(-z/z_{\rm flux})$ 
of the phase space density and the flux, with: 
\begin{equation}
z_{\rm p.s.}=d_0\, \frac{\bar v}{\Delta v_0}\; 
\frac{e^\eta}{\pi^{3/2}(\eta+\tilde{S}(\eta)-2)S(\eta)}
\qquad \qquad
z_{\rm flux}=d_0\, \frac{\bar v}{\Delta v_0}\;
 \frac{e^\eta}{\pi^{3/2}S(\eta)}
\label{lengths}
\end{equation}
This explains why the results in figs. \ref{DATAa},\ref{DATAb} exhibit a quasi-exponential
variation with $z$.
For $\eta=5$, we get
$z_{\rm p.s.}\simeq 100\; d_0\, {\bar v}/{\Delta v_0}$ and 
$z_{\rm flux}\simeq 360\; d_0\, {\bar v}/{\Delta v_0}$. The factor $3.6$
between these two lengths suggests that a phase space gain of 7 orders of
magnitude can be achieved with a flux reduction by less than 2 orders of 
magnitude.
This is confirmed by the numerical solution of the hydrodynamical equations
Eq.(\ref{HYnum1}-\ref{HYnum3}): the required length is $7600\; d_0$, on the
order of $11$ meters for the parameters of section \ref{MODEL}; 
it is accompanied by a decrease of temperature by a factor 4000 and of flux
by a factor 90.

It is worth noting that the collision rate $\gamma_{\rm coll}(z)$ does not 
vary much for $z \ll |z_c|$.
Indeed, we have: 
$$
\gamma_{\rm coll}(z)=\gamma_{\rm coll}(0) \left( 1-\frac{z}{z_c} \right)^{\alpha_c
-\frac{\beta_c}{2}}=\gamma_{\rm coll}(0)\frac{z_c}{z_c-z} \quad.
$$ 
The sign of $z_c$ is the same as the sign of $\eta+\tilde S(\eta)-6$, which vanishes
for $\eta\simeq 5.4$.
Therefore, the collision
rate $\gamma_{\rm coll}$
increases with $z$ if  $\eta >5.4 $ and decreases if $\eta <5.4 $. For the
particular value $\eta= 5.4$, 
the collision rate is constant, as well as the loss rates for
particles $\Gamma(z)$ and energy $\Gamma_\epsilon(z)$; the 
quantities  $\rho_{\rm lin}(z)$ and $k_BT(z)$ then have 
an exponential variation for any $z$, since $z_c,\alpha_c,\beta_c
\rightarrow \infty$.

\subsection{ 2D versus 1D evaporation scheme} \label{2D1D}

The experimental setup based on radio-frequency evaporation
and discussed in section \ref{MODEL}, corresponds to a 2D 
evaporation scheme while the analytical treatment of section \ref{APPROX}
is based on a 1D evaporation. 
One does not expect the two situations to be equivalent.
More precisely, 
since the $x$-$y$ degrees of freedom  are not mixed
by the collisionless motion in the axi-symmetric potential,
the 1D evaporation is expected to be less efficient 
than a 2D evaporation scheme: 
a particle may emerge from an elastic collision with a large transverse
kinetic energy along the $y$-axis, without being evaporated in the
1D evaporative scheme.

We have checked with our Monte-Carlo simulation that 2D evaporation is 
more  efficient indeed than 1D evaporation. 
We have first used a crude model of 2D evaporation, with the truncation
of Eq.(\ref{ANS}) replaced by $Y(\epsilon_c(z)-\epsilon_{\rm tot}+k_B T(z)/2)$, where
$\epsilon_{\rm tot}=\epsilon_\perp +(p_z-\bar{p}(z))^2/2m$ is
the total energy in the frame moving at velocity $\bar{p}(z)/m$. 
This truncation assumes in particular quasi-ergodicity in the $x-y$ plane.
Within this model we perform the same optimization as in fig. \ref{eta};
to achieve a gain in phase space density by 7 orders of magnitude the optimal
$\eta$ is now equal to 6.
We have then run the Monte-Carlo simulation for this value of $\eta$.
The gain of a factor 500 in phase space density, which was obtained for a length
$2500 \; d_0$ for 1D evaporation (see fig. \ref{DATAb}), is reached now
for a length $900 \; d_0$; the relative variation of the flux (reduction by a factor
$\sim 6$) is similar. The variations with $z$ of the phase space density
and the flux are quasi-exponential. Note that the predictions of the crude model
are in amazingly good agreement with the Monte-Carlo results.

To conclude the length of the guide
needed to achieve a specified gain of the phase-space 
density is a factor of $\sim 3$ smaller for 2D
than for 1D evaporation. E.\ g.\ the 11 meters of evaporation length necessary in 1D
for the experimental conditions of \S \ref{MODEL}
are now reduced to 4 meters.

\section{Perspectives}\label{sec_per}

For a typical experimental source of cold atoms,
the length needed for a phase-space increase of
seven orders of magnitude -- which should bring the system close 
to the degeneracy point -- is of the order of a few meters.
The possibility of quantum degeneracy in such a system
raises interesting questions. As it is well
known\cite{Huang}, there is no Bose-Einstein 
condensation in the thermodynamic limit
in a 1D geometry, obtained here \cite{FOOT}
in setting
$N,L \rightarrow \infty$, while keeping a constant
linear density $\rho_{\rm lin}=N/L$, a constant temperature
$T$ and a constant transverse oscillation frequency $\Omega_\bot$.
Therefore we do not expect 
a macroscopic occupation of a 
single quantum state of the longitudinal motion.

In order to get more insight in the output of this system, 
we assume that the
transverse extension of the thermal component of the beam 
$R_\bot=\sqrt{k_B T/ (m\Omega_\bot^2)}$ is much larger than
the thermal wavelength $\lambda=h/\sqrt{2\pi m k_B T}$ 
(that is $k_B T \gg \hbar \Omega_\bot$).
In this case, elastic collisions ensure
that thermodynamic equilibrium is reached in the frame moving
with the mean velocity of the gas. 
In this frame, we 
describe the properties of the system using as an approximation
the grand canonical Bose-Einstein 
distribution for an ideal gas. 
The gas being
at temperature $T$, we obtain the maximal linear density
that can be put in states corresponding to an excited transverse
motion 
\begin{equation}
\rho_{\rm lin}^{(c)}= \frac{1}{\lambda} \; \zeta(5/2)\;
\left(\frac{k_B T}{\hbar \Omega_\bot} \right)^2
\end{equation}
with $\zeta(5/2) \simeq 1.34$.
When one increases the linear density above this critical value
(which corresponds to a spatial density on axis 
larger than  $n^{(c)}=\zeta(3/2)\lambda^{-3}$),
the transverse degrees of freedom undergo a Bose-Einstein
condensation\cite{Ketterle96}, that is  
atoms start accumulating in states corresponding to the ground
transverse motion (see fig. \ref{BOSE}). 
By convention, we set the energy of this ground state to zero.
Since the chemical potential $\mu$
then satisfies $|\mu|\ll \hbar \Omega_\bot \ll k_B T$,
the longitudinal 
momentum distribution for these atoms can be approximated as:
\begin{equation}
n(p)=\frac{1}{h}\; \frac{1}{\exp\left[(p^2/(2m)-\mu)/k_BT\right]-1} \simeq
\frac{h}{\pi\lambda^2}\;\frac{1}{p^2+p_c^2} 
\label{POP}
\end{equation} 
that is a Lorentzian distribution of half width $p_c=[-2m\mu]^{1/2}$.
Such a distribution leads to a spatial correlation length
of the gas along $z$ given by $\xi_c=\hbar/p_c$.
By integrating $n(p)$ over $p$, we can relate $\mu$ and therefore
$\xi_c$ to the linear density of atoms in the transverse
ground state $\rho_{\rm lin}-\rho_{\rm lin}^{(c)}$:
\begin{equation}
\xi_c = \frac{\lambda^2}{2\pi}
(\rho_{\rm lin} - \rho_{\rm lin}^{(c)})
=\frac{1}{2} \zeta(3/2)
{\hbar \over m \Omega_\bot \lambda} {n -n^{(c)} \over 
n^{(c)}}
\label{xi}
\end{equation} 
where  $n$ is the 3D on-axis density.
This correlation length can be much larger than the thermal wavelength.
Thus, for $\rho_{\rm lin} > \rho_{\rm lin}^{(c)}$,
the output of the system can be viewed as propagating 
independent trains of matter waves, each of which having
a length of the order of $\xi_c$ and  containing on average $\left[
\lambda(\rho_{\rm lin}-\rho_{\rm lin}^{(c)} )\right]^2/(2\pi)$ atoms. 

 To conclude, we have presented in this paper the principles of the
evaporative cooling of an atomic beam. In the classical regime, 
where the mean interparticle distance is much larger than the thermal
wavelength $\lambda$, we have evaluated the characteristic lengths 
for the flux and the phase space density variations.
We have also outlined briefly the coherence properties of the beam once quantum
degeneracy is reached.
A more detailed characterization of these coherence properties,
including the effect of interactions between  particles,
will be the subject of a future work.

\section*{Acknowledgments}
We thank F. Chevy, C. Cohen-Tannoudji, D. Gu\'ery-Odelin,
 K. Madison,
C. Salomon, 
and G. Shlyapnikov 
for several helpful discussions. 
E.M. also acknowledges useful discussions
with A. Aspect and P. Bouyer.
Laboratoire Kastler Brossel is a {\it unit\'e de recherche du CNRS}, 
associated with Ecole Normale Sup\'erieure and Universit\'e Pierre et Marie
Curie. A.M. and I. C. are with the Istituto Nazionale per la Fisica della
Materia. This work was partially supported by CNRS, Coll\`{e}ge de France,
DRET, DRED and EC (TMR network ERB FMRX-CT96-0002).

\newpage

\section{Appendix}
In this appendix we derive explicit expressions for the loss rates for particles 
$\Gamma$
and for energy $\Gamma_\epsilon$ in Eqs. (\ref{HY1},\ref{HY2},\ref{HY3}).
Integrating (\ref{BOLTZ}) over $x$, $y$, and $\bf p$ to derive (\ref{HY1}), 
we obtain:
\begin{eqnarray}
\Gamma(z) \, \rho_{\rm lin}(z) &=& \frac{\sigma}{ \pi m^2 }
\int dx\;dy \int d^3 p \int d^3 p_2 \int d^3 p_3 \int d^3 p_4 
\; \left(f({\bf r}, {\bf p})\, f({\bf r}, {\bf p}_2)
- f({\bf r}, {\bf p}_3)\, f({\bf r}, {\bf p}_4) \right)
\nonumber \\
&& \delta ({\bs p} + {\bs p}_2 - {\bs p}_3 - {\bs p}_4)
\; \delta \left(\frac{p^2}{2m}+\frac{p_2^2}{2m}- \frac{p_3^2}{2m}- \frac{p_4^2}{2m}
\right).
\label{TC2}
\end{eqnarray}
In this integral, the phase space cell ($\bs r, \bs p$) is the variable
of integration of the left hand side of  (\ref{BOLTZ}) and it is always
in the trappable domain defined by:
\begin{equation}
\frac{p_x^2}{2m}+\frac{1}{2}m \Omega_\bot^2 x^2 < \epsilon_c(z)
\end{equation} 
For the first part of
the integral, representing the collision ${\bf p} + {\bf p}_2 \rightarrow
{\bf p}_3 + {\bf p}_4$ and 
involving $f({\bf r}, {\bf p})\, f({\bf r}, {\bf p}_2)$,
the cell (${\bs r}, {\bs p}_2$) is also in the trappable domain, while the two
cells (${\bs r}, {\bs p}_3$) and (${\bs r}, {\bs p}_4$) may be either in or out
of the trappable domain.
For the second part of the
integral, representing the collision ${\bf p}_3 + {\bf p_4} \rightarrow
{\bf p} + {\bf p_2}$ and 
involving $f({\bf r}, {\bf p}_3)\, f({\bf r}, {\bf p}_4)$,
the two cells (${\bs r}, {\bs p}_3$) and (${\bs r}, {\bs p}_4$) are in the
trappable domain, while the cell (${\bs r}, {\bs p}_2$) may be either in or out
of the trappable domain. 

We now rearrange the second part of this integral by exchanging the role of
$({\bf p},{\bf p}_2)$ and $({\bf p}_3,{\bf p}_4)$. After cancellation of various
terms, we are left with: 
\begin{eqnarray}
\Gamma(z) \; \rho_{\rm lin}(z) &=& \frac{\sigma}{ \pi m^2 }
\int dx\;dy \int d^3 p_1 \int d^3 p_2 \int d^3 p_3 \int d^3 p_4 
\; f({\bf r}, {\bf p_1})\, f({\bf r}, {\bf p}_2)
\nonumber \\
&& \delta ({\bs p}_1 + {\bs p}_2 - {\bs p}_3 - {\bs p}_4)
\; \delta \left(\frac{p_1^2}{2m}+\frac{p_2^2}{2m}- \frac{p_3^2}{2m}- \frac{p_4^2}{2m}
\right).
\label{TC3}
\end{eqnarray}
where (${\bs r}, {\bs p}_1$) and (${\bs r}, {\bs p}_2$) are in the trappable
domain, (${\bs r}, {\bs p}_3$) is out of the trappable domain, and
(${\bs r}, {\bs p}_4$) is either in or out of the trappable domain.

We now put 
%\begin{eqnarray*}
\begin{equation}
{\bf P}=\frac{{\bf p}_1 + {\bf p}_2}{2} 
%&\qquad &
\qquad
{\bf q}=\frac{{\bf p}_1 - {\bf p}_2}{2} 
%\\
\qquad
{\bf P'}=\frac{{\bf p}_3 + {\bf p}_4}{2} 
%&\qquad &
\qquad
{\bf q'}=\frac{{\bf p}_3 - {\bf p}_4}{2} \quad.
\end{equation}
%\end{eqnarray*}
The $\delta$-distributions entering into (\ref{TC3}) impose ${\bf P}={\bf P'}$
and $|{\bf q}|=|{\bf q'}|$. We perform some rearrangements, using as 
integration variables for the vectors ${\bf q}$ and 
${\bf q'}$ the coordinates $q_x,q_x'$ along $x$, the moduli
$q,q'$ and the azimuthal angles around $x$-axis. We integrate over those
angles and we split the
integration domain into a part where only particle 3 escapes
and a part where both 3 and 4 escape.
We then obtain: 
\begin{eqnarray}
{\Gamma}(z)\; \rho_{\rm lin} (z)& = &
64 \pi \frac{\sigma}{m} f_0^2(z) 
\int_{-x_c}^{x_c} dx 
\int_0^Q d P_x 
\int_{-\infty}^{\infty} dy 
\int_{-\infty}^{\infty} dP_y 
\int_{-\infty}^{\infty} dP_z
\; 
e^{-2(U_x+U_y)/k_B T}\; 
e^{-P^2/mk_B T} 
\nonumber \\
& \times & 
\int_0^{Q-P_x} dq_x \; 
\left\{ 
\int_{Q-P_x}^{Q+P_x} dq \; q\; e^{-q^2/mk_B T}
\int_{Q-P_x}^q dq^{\prime}_x \right.
\nonumber \\
&& \qquad \qquad \qquad + \left.
\int_{Q+P_x}^{\infty} dq \; q \; e^{-q^2/mk_B T}
 \left[ 
\int_{Q-P_x}^{Q+P_x} dq_x^{\prime} + 
2\int_{Q+P_x}^q dq_x^{\prime} \right] \right\} 
\label{TC4}
\end{eqnarray}
where $P=|{\bf P}|$ and $q=|{\bf q}|$. The length
$x_c(z)$ is the positive solution of $m \Omega_{\bot}^2 x_c^2(z) =2 \epsilon_c(z)$,
and $Q(x,z)$ is the local escape momentum given by 
$Q^2/2m=\epsilon_c(z)-U_x(x)$.
The expression (\ref{TC4}) is the direct transcription
of Eq. (26) of \cite{amsterdam2}, to the situation considered in the present
paper. 

The integrations over $y,P_y,P_z$ are immediate, as well as for $q_x$. After an integration by part over $q$, we get:
\begin{equation}
{\Gamma}\, \rho_{\rm lin}  = 
16 \, \pi^3 \,\sigma\, (k_BT)^{9/2}\, m^{3/2}\, \Omega_\bot^{-2}\, f_0^2\, 
\int_{-\sqrt{2\eta}}^{\sqrt{2\eta}} dw \; e^{-w^2} \int_0^a du \; (a-u)\;
e^{-u^2}\;({\rm erfc}(a-u)+{\rm erfc}(a+u))
\end{equation}
where we have put $w=x/R_\bot$, $a=Q/\sqrt{mkT}=\sqrt{2\eta-w^2}$, $u=P_x/\sqrt{mkT}$
and:
$$
{\rm erfc}(u)=\frac{2}{\sqrt{\pi}}\; \int_u^\infty dv \; e^{-v^2}
\quad.
$$
Using finally the relation (\ref{NORM}) to express $f_0$ in terms of the
linear density $\rho_{\rm lin}$, we arrive to (\ref{LOSS1}) with:
\begin{equation}
S(\eta)=\frac{1}{2\pi^2}\; \frac{e^\eta}{(1-e^{-\eta})^2}\;
\int_{-\sqrt{2\eta}}^{\sqrt{2\eta}} dw \; e^{-w^2} \int_0^a du \; (a-u)\;
e^{-u^2}\;({\rm erfc}(a-u)+{\rm erfc}(a+u)).
\label{OUF1}
\end{equation}

The calculation of $\Gamma_\epsilon$ proceeds along the same line.
We multiply (\ref{BOLTZ}) by $\epsilon_x+\epsilon_y+p_z^2/2m$ and integrate
over $x$, $y$, and ${\bf p}$. Using the same rearrangement as above, we 
get:
\begin{eqnarray}
{\Gamma_\epsilon}\,k_B T\, \rho_{\rm lin} & = &
64 \pi \frac{\sigma}{m} f_0^2 
\int_{-x_c}^{x_c} dx 
\int_0^Q d P_x 
\int_{-\infty}^{\infty} dy 
\int_{-\infty}^{\infty} dP_y 
\int_{-\infty}^{\infty} dP_z
\; 
e^{-2(U_x+U_y)/k_B T}\; 
e^{-P^2/mk_B T} 
\nonumber \\
& \times & 
\int_0^{Q-P_x} dq_x \; 
\left\{ 
\int_{Q-P_x}^{Q+P_x} dq \; q\; e^{-q^2/mk_B T}
\int_{Q-P_x}^q E_3\; dq^{\prime}_x \right.
\nonumber \\
&& \qquad \qquad \qquad + \left.
\int_{Q+P_x}^{\infty} dq \; q \; e^{-q^2/mk_B T}
 \left[ 
\int_{Q-P_x}^{Q+P_x} E_3\; dq_x^{\prime} + 
\int_{Q+P_x}^q (E_3+E_4)\; dq_x^{\prime} \right] \right\} 
\label{TC5}
\end{eqnarray}
where:
\begin{eqnarray*}
E_3 &=& \frac{p_3^2}{2m}+U_x+U_y=\frac{({\bf P'}+{\bf q'})^2}{2m} + U_x+U_y \\
E_4 &=& \frac{p_4^2}{2m}+U_x+U_y=\frac{({\bf P'}-{\bf q'})^2}{2m} + U_x+U_y.
\end{eqnarray*}
The expression (\ref{TC5}) is also the direct transposition of 
Eq. (26) of \cite{amsterdam2}.
After integration and changes of variables similar to the ones given above
for the calculation of $S(\eta)$, we reach (\ref{LOSS2}) where 
\begin{eqnarray}
\tilde{S}(\eta)&=&\frac{1}{4\pi^2\;S(\eta)}\; \frac{e^\eta}{(1-e^{-\eta})^2}\;
\int_{-\sqrt{2\eta}}^{\sqrt{2\eta}} dw \; e^{-w^2} \int_0^a du \; (a-u)\;
e^{-u^2} \nonumber \\
&& \times \left\{
\frac{a+u}{\sqrt \pi}e^{-(a-u)^2}+\frac{a-u}{\sqrt \pi}e^{-(a+u)^2}
-(a^2-u^2)({\rm erfc}(a-u)+{\rm erfc}(a+u))
\right\} \quad.
\label{OUF2}
\end{eqnarray} 
with $a=\sqrt{2 \eta -w^2}$ as in (\ref{OUF1}).
\newpage

\begin{figure}[ht]
\caption{An atomic beam propagates 
in a transverse magnetic guide.
Evaporation eliminates particles whose 
transverse coordinate exceeds some adjustable
value $\Lambda_{\rm evap}(z)$. The emerging beam 
is colder and it has a larger phase-space density than the input beam.
 \label{SHEME}}
\end{figure}

\begin{figure}[ht]
\caption{Variations with $\eta$ of the dimensionless parameters $S$ 
multiplied by 10 for clarity (full line)
 and $\tilde{S}$ (dashed line).
The analytical expressions for these parameters are derived in the appendix.}
\label{fig:S}
\end{figure}

\begin{figure}[ht]
\caption{Longitudinal velocity as a
function of the position $z$ measured in units of
the initial mean free path $d_0=\sqrt{\pi}/(2n_0 \sigma)$.
The unit for velocity is the initial velocity spread $\Delta v_0$.
The continuous line corresponds to the numerical solution
of the 1D hydrodynamic equation for $\eta=5$. 
The markers indicate the results of
the Monte-Carlo simulation. 
\label{VELO}}
\end{figure}

\begin{figure}[ht]
\caption{Flux as a
function of the position $z$ measured in units of
the initial mean free path  $d_0=\sqrt{\pi}/(2n_0 \sigma)$. 
The unit of flux is the initial value at $z=0$.
The continuous line corresponds to the numerical solution
of the 1D hydrodynamic equation for $\eta=5$. The markers indicate the results of
the Monte-Carlo simulation. 
 \label{DATAa}}
\end{figure}

\begin{figure}[ht]
\caption{Gain in phase-space density as a
function of the position $z$ measured in units of
the initial mean free path  $d_0=\sqrt{\pi}/(2n_0 \sigma)$.
The continuous line corresponds to the numerical solution  
of the 1D hydrodynamic equation with $\eta=5$. The markers indicate the results of
the Monte-Carlo simulation.
 \label{DATAb}} 
\end{figure}

\begin{figure}[ht]
\caption{Gain in phase space density as a function of the position
$z$ expressed in units 
of the initial mean free path  $d_0=\sqrt{\pi}/(2n_0 \sigma)$. These curves are the 
numerical solutions
of Eqs. (\ref{HYnum1}-\ref{HYnum3}) in the case where $\eta$ is fixed.}
\label{eta}
\end{figure}

\begin{figure}[ht]
\caption{For an ideal Bose gas 
fraction of atoms in the transverse ground state, as a function 
of the linear density.
This figure
has been obtained for $k_B T =20\; \hbar \Omega_\bot$, so that 
$\rho_{\rm lin}^{(c)} \simeq 536\;\lambda^{-1} $.} 
\label{BOSE}
\end{figure}


\begin{thebibliography}{99}
\bibitem{gravity} To compensate for the deceleration of the guided atoms 
caused by the variation of $B_0$
with $z$, one can tilt slightly the guide with respect to the horizontal position.
Alternatively the corresponding extra variation of $\bar p$ can be 
incorporated in our treatment.
\bibitem{remarque} Quasi-ergodicity in the 2D evaporation scheme could
be obtained by introducing a slight non-axi-symmetric potential, e.g.\
with a far detuned laser beam. A relative oscillation frequency difference
between $x$ and $y$ axis $\Delta \Omega/\Omega_\perp
\sim \pi \gamma_{\rm coll}/\Omega_\perp\sim 6\%$ 
is sufficient to mix the $x$
and $y$ motions between two collisions.
If this is achieved a formulation of 2D evaporation in terms of
the transverse energy $\epsilon_\perp$ is possible. Note however
that the calculations are in this case much more involved than for
1D or 3D evaporation.
\bibitem{SPIL} The approximate exponential variations found in
section \ref{sec_disc} allow an estimation
of the loss rate $\Gamma_{\rm sp}$ of particles due to spilling.
One finds $\Gamma_{\rm sp}/\Gamma
\simeq 2\eta(\eta+\tilde{S}-1)/(5 e^\eta-5)$ which is on the order
of $6\%$ for $\eta=5$.
\bibitem{FOOT}
This should not be confused with the 3D thermodynamic limit 
for this problem, which is obtained by letting $N,L \rightarrow \infty$,
$\Omega_\bot \rightarrow 0$, while keeping a constant on-axis density
and a constant product $L\Omega_\bot$.
\end{thebibliography}
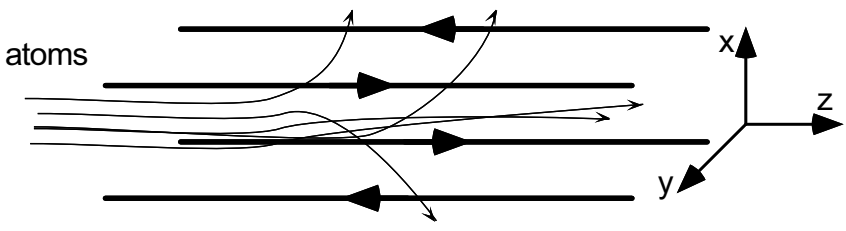
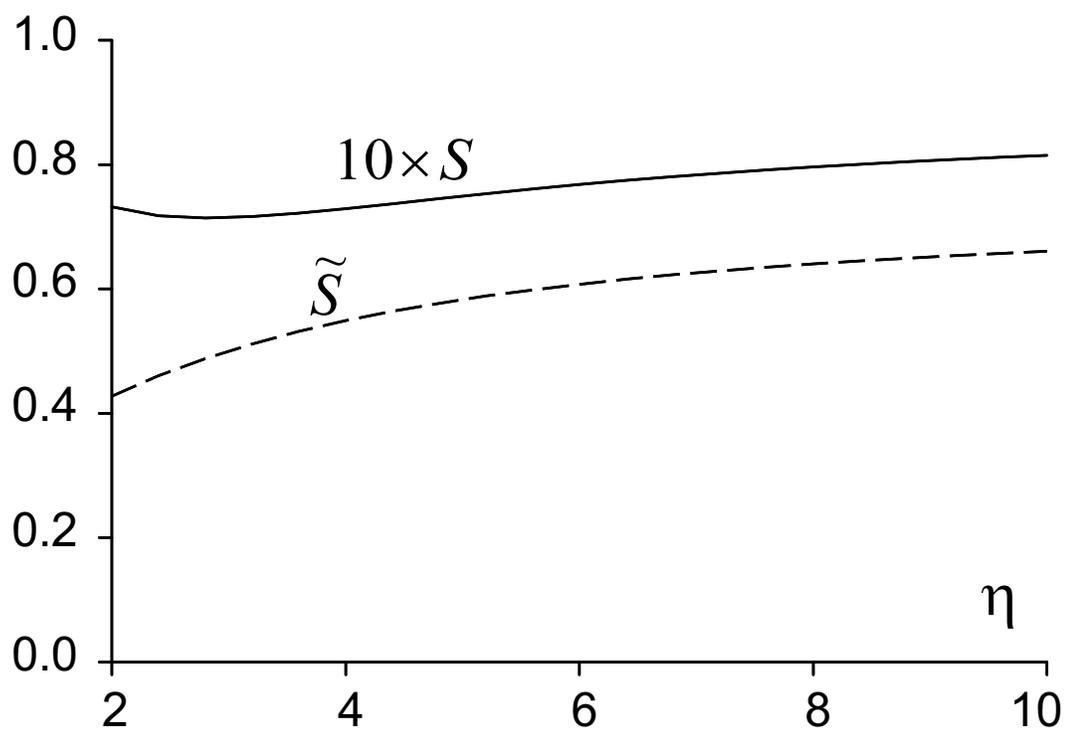
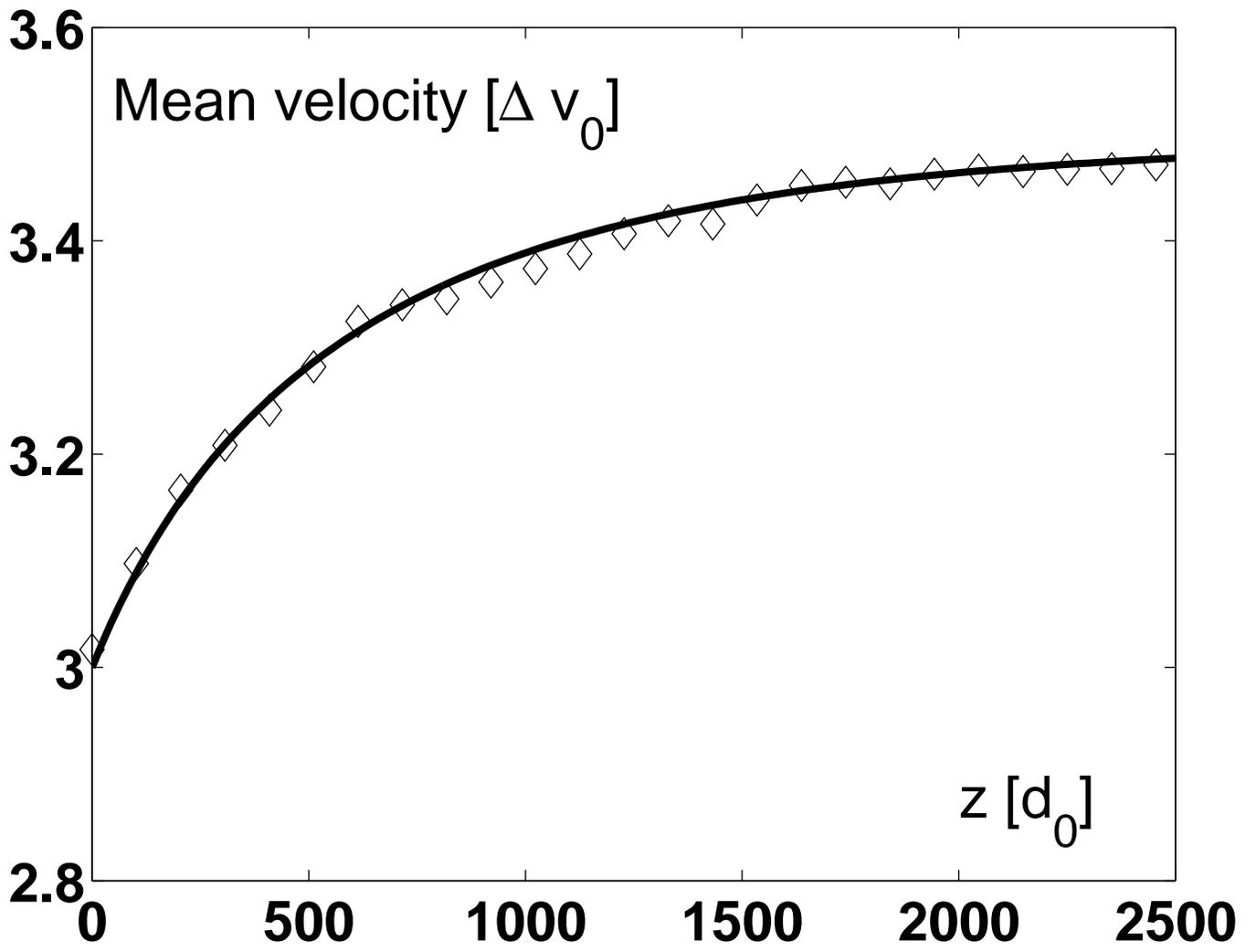
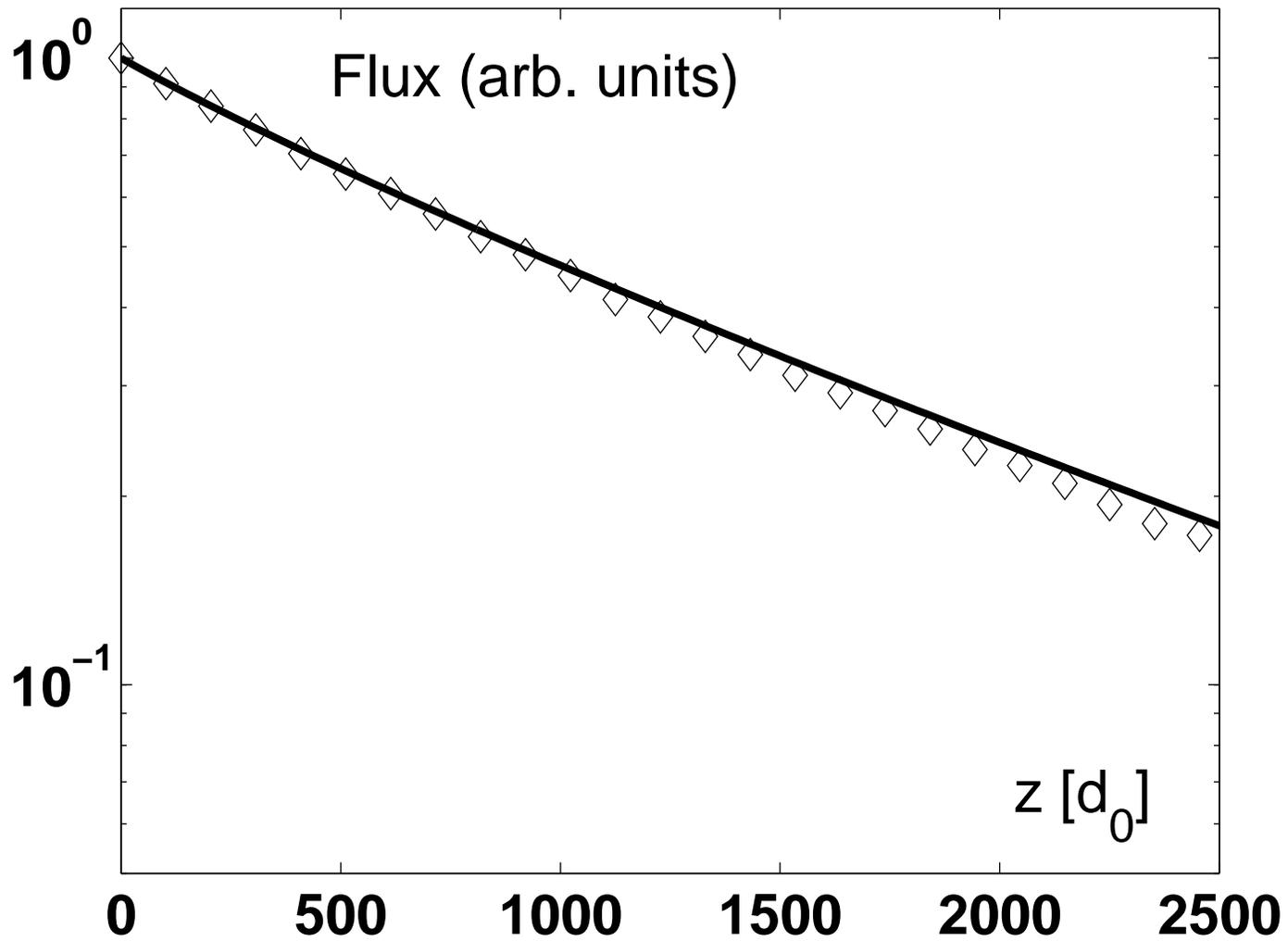
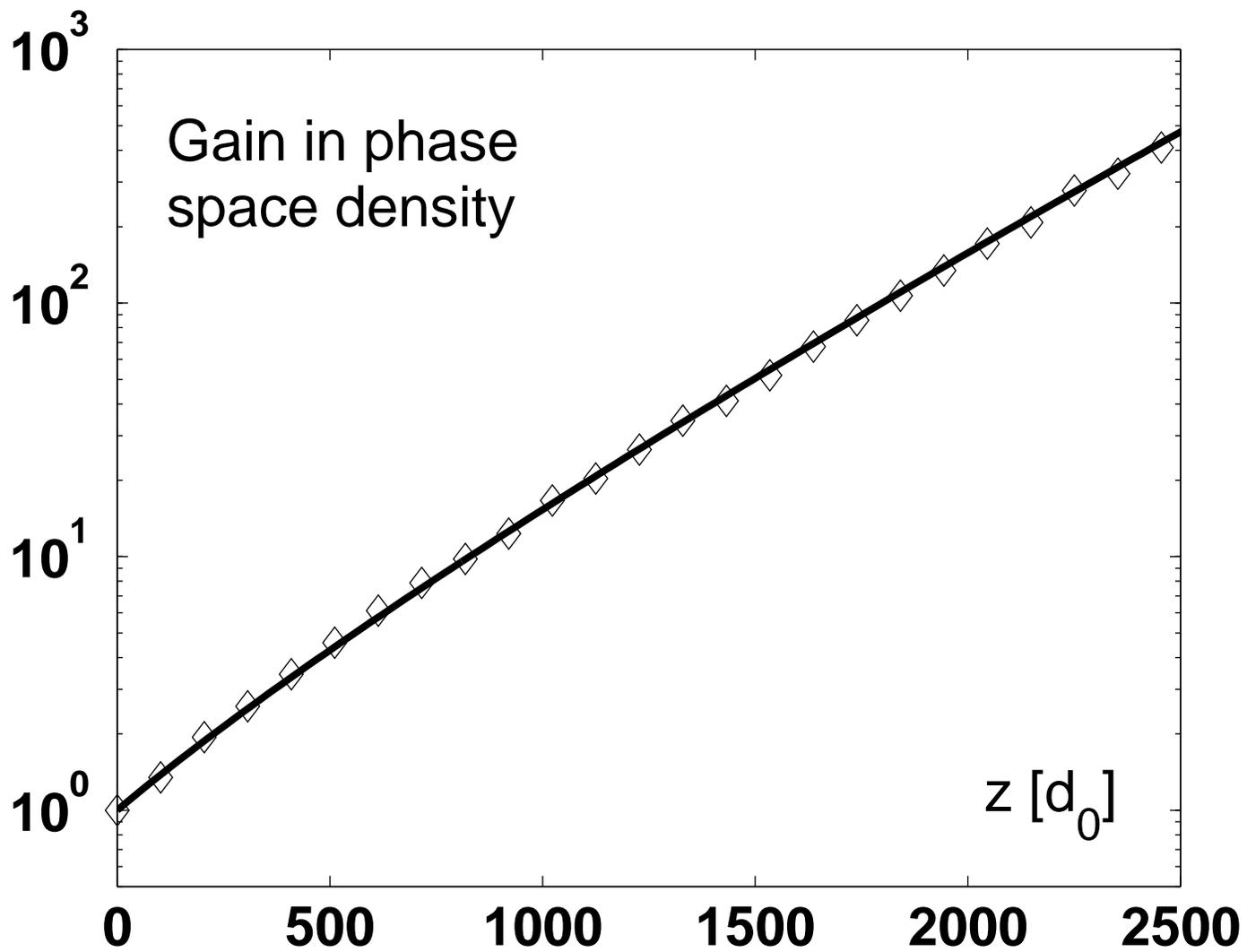
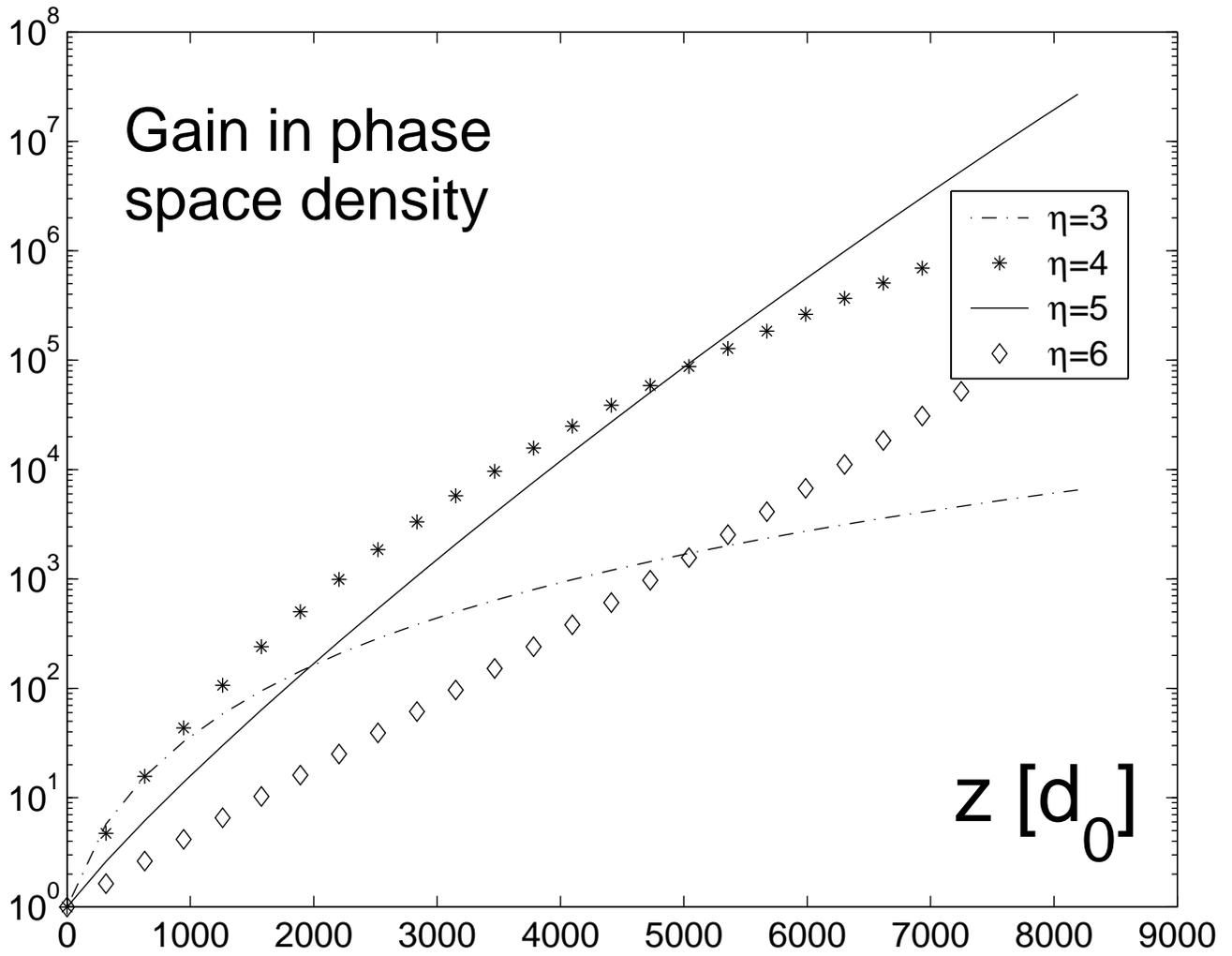
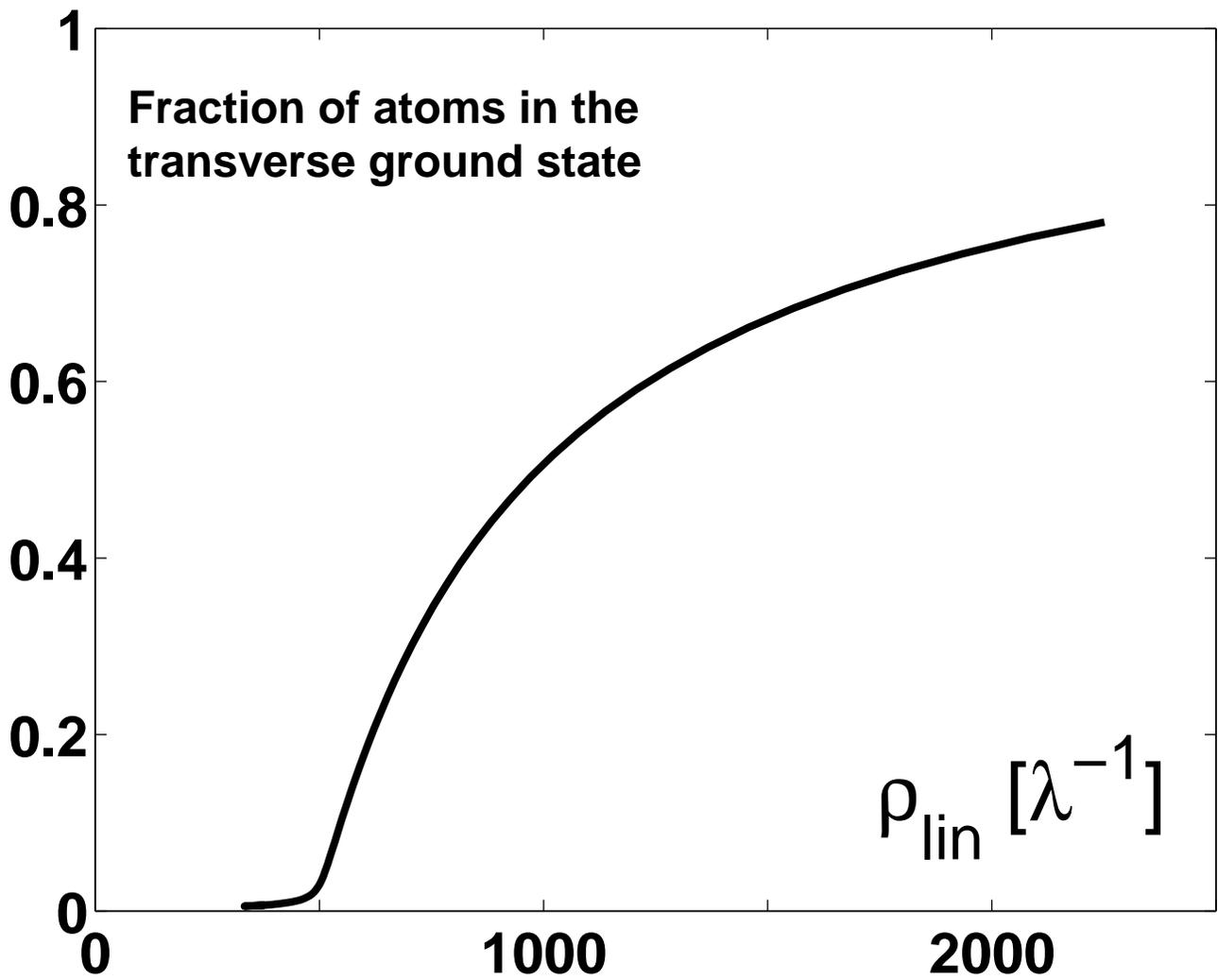
\end{document}